\documentclass[letterpaper]{article}
\usepackage{aaai}
\usepackage{times}
\usepackage{helvet}
\usepackage{courier}

\usepackage{times}

\usepackage{amsmath}
\usepackage{amsfonts}
\usepackage{amssymb}
\usepackage{cases}
\usepackage{pst-tree}
\usepackage{multirow}
\usepackage{footnote}
\usepackage{tabularx}
\usepackage{fixfoot}
\usepackage{lscape}
\usepackage{arydshln}
\usepackage{MnSymbol}
\usepackage{eurosym}
\usepackage{algorithm}
\usepackage[noend]{algorithmic}
\usepackage[chatter]{rotating}

\newtheorem{thm}{Theorem}

\newtheorem{lem}[thm]{Lemma}
\newtheorem{prop}[thm]{Proposition}
\newtheorem{defn}[thm]{Definition}
\newtheorem{exam}[thm]{Example}

\begin{document}
\title{Efficient evolutionary dynamics with extensive--form games}
\author{%Paper ID 269
Nicola Gatti \and Fabio Panozzo \and Marcello Restelli\\
Politecnico di Milano \\
Piazza Leonardo da Vinci 32 \\
Milano, Italy \\
\{ngatti, panozzo, restelli\}@elet.polimi.it\\
}
\maketitle
\begin{abstract}
Evolutionary game theory combines game theory and dynamical systems and is customarily adopted to describe evolutionary dynamics in multi--agent systems. In particular, it has been proven to be a successful tool to describe multi--agent learning dynamics. To the best of our knowledge, we provide in this paper the first replicator dynamics applicable to the \emph{sequence form} of an extensive--form game, allowing an exponential reduction of time and space w.r.t. the currently adopted replicator dynamics for normal form. Furthermore, our replicator dynamics is realization equivalent to the standard replicator dynamics for normal form. We prove our results for both discrete--time and continuous--time cases. Finally, we extend standard tools to study the stability of a strategy profile to our replicator dynamics.
\end{abstract}

\section{Introduction}

Game theory provides the most elegant tools to model strategic interaction situations among rational agents. These situations are customarily modeled as \emph{games}~\cite{fudenberg1991} in which the \emph{mechanism} describes the rules and \emph{strategies} describe the behavior of the agents. Furthermore, game theory provides a number of \emph{solution concepts}. The central one is \emph{Nash equilibrium}. Game theory assumes agents to be rational and describes ``static'' equilibrium states. Evolutionary game theory~\cite{cressman2003} drops the assumption of rationality and assumes agents to be adaptive in the attempt to describe dynamics of evolving populations. Interestingly, there are strict relations between game theory solution concepts and evolutionary game theory steady states, e.g., Nash equilibria are steady states. Evolutionary game theory is commonly adopted to study economic evolving populations~\cite{parsons2007} and artificial multi--agent systems, e.g., for describing multi--agent learning dynamics~\cite{tuyls2006,tuyls2007,tuyls2008} and as heuristics in algorithms~\cite{security2010}. In this paper, we develop efficient techniques for evolutionary dynamics with extensive--form games.

Extensive--form games are a very important class of games. They provide a richer representation than strategic--form games, the sequential structure of decision--making being described explicitly and each agent being allowed to be free to change her mind as events unfold. The study of extensive--form games is carried out by translating the game by means of tabular representations~\cite{shoham-book}. The most common is the \emph{normal form}. Its advantage is that all the techniques applicable to strategic--form games can be adopted also with this representation. However, the size of normal form grows exponentially with the size of the game tree, thus being impractical. The \emph{agent form} is an alternative representation whose size is linear in the size of the game tree, but it makes, even with two agents, each agent's best--response problem highly non--linear. To circumvent these issues, \emph{sequence form} was proposed~\cite{stengel1996}. This form is linear in the size of the game tree and does not introduce non--linearities in the best--response problem. On the other hand, standard techniques for strategic--form games cannot be adopted with such representation, e.g.~\cite{lemke1964}, thus requiring alternative \emph{ad hoc} techniques, e.g.~\cite{lemke1978}. In addition, sequence form is more expressive than normal form. For instance, working with sequence form it is possible to find Nash--equilibrium refinements for extensive--form games---perfection based Nash equilibria and sequential equilibrium~\cite{quasiperfect,GattiIulianoAAAI2011}---while it is not possible with normal form.

To the best of our knowledge, there is no result dealing with the adoption of evolutionary game theory tools with sequence form for the study of extensive--form games, all the known results working with the normal form~\cite{cressman2003}. In this paper, we originally explore this topic, providing the following main contributions.
\vspace{-0.15cm}
\begin{itemize}
\item We show that the standard replicator dynamics for normal form cannot be adopted with the sequence form, the strategies produced by replication not being well--defined sequence--form strategies.
\vspace{-0.15cm}
\item We design an \emph{ad hoc} version of the discrete--time replicator dynamics for sequence form and we show that it is sound, the strategies produced by replication being well--defined sequence--form strategies.
\vspace{-0.15cm}
\item We show that our replicator dynamics is realization equivalent to the standard discrete--time replicator dynamics for normal form and therefore that the two replicator dynamics evolve in the same way.
\vspace{-0.15cm}
\item We extend our discrete--time replicator dynamics to the continuous--time case, showing that the same properties are satisfied and extending standard tools to study the stability of the strategies to our replicator.
\end{itemize}

\vspace{-0.3cm}
\section{Game theoretical preliminaries}

\textbf{Extensive--form game definition}. A \emph{perfect--information} extensive--form game~\cite{fudenberg1991} is a tuple $(N, A, V, T, \iota, \rho, \chi, \mathbf u)$, where: $N$ is the set of agents ($i\in N$ denotes a generic agent), $A$ is the set of actions ($A_i\subseteq A$ denotes the set of actions of agent~$i$ and $a\in A$ denotes a generic action), $V$ is the set of  decision nodes ($V_i\subseteq V$ denotes the set of decision nodes of~$i$), $T$ is the set of terminal nodes ($w\in V\cup T$ denotes a generic node and $w_0$ is root node), $\iota:V\rightarrow N$ returns the agent that acts at a given decision node, $\rho:V\rightarrow \wp(A)$ returns the actions available to agent $\iota(w)$ at $w$, $\chi:V\times A \rightarrow V\cup T$ assigns the next (decision or terminal) node to each pair $\langle w,a \rangle$ where $a$ is available at~$w$, and $\mathbf u=(u_1,\ldots,u_{|N|})$ is the set of agents' utility functions $u_i:T\rightarrow \mathbb{R}$.  Games with \emph{imperfect information} extend those with perfect information, allowing one to capture situations in which some agents cannot observe some actions undertaken by other agents. We denote by $V_{i,h}$ the $h$--th \emph{information set} of agent~$i$. An information set is a  set of decision nodes such that when an agent plays at one of such nodes she cannot distinguish the node in which she is playing. For the sake of simplicity, we assume that every information set has a different index~$h$, thus we can univocally identify an information set by~$h$. Furthermore, since the available actions at all nodes~$w$ belonging to the same information set $h$ are the same, with abuse of notation, we write $\rho(h)$ in place of $\rho(w)$ with $w\in V_{i,h}$. An imperfect--information game is a tuple $(N, A, V, T, \iota, \rho, \chi, \mathbf u, H)$ where $(N, A, V, T, \iota, \rho, \chi, \mathbf u)$ is a perfect--information game and $H = (H_1,\ldots,H_{|N|})$ induces a partition $V_i = \bigcup_{h\in H_i} V_{i,h}$ such that for all $w,w' \in V_{i,h}$ we have $\rho(w) = \rho(w')$.  We focus on games with \emph{perfect recall} where each agent recalls all the own previous actions and the ones of the opponents~\cite{fudenberg1991}.
 
\begin{figure}[htb]
\vspace{-0.5cm}
\centering
%\documentclass[12pt]{article}
%
%\usepackage{pifont,pst-optic}
%
%\begin{document}
%
%\thispagestyle{empty}

\begin{pspicture}(-1,0.5)(3.5,-4.0)
\scalebox{0.75}{

\psdots[linecolor=black](0.5,0)(-1,-1.2)(2,-1.2)(1,-2.4)(3,-2.4)(1.5,-3.6)(3.5,-3.6)(0.5,-3.6)(2.5,-3.6)

\psline[linecolor=black,linewidth=0.5pt]{-}(0.5,0)(-1,-1.2)
\psline[linecolor=black,linewidth=0.5pt]{-}(0.5,0)(2,-1.2)
\psline[linecolor=black,linewidth=0.5pt]{-}(2,-1.2)(1,-2.4)
\psline[linecolor=black,linewidth=0.5pt]{-}(2,-1.2)(3,-2.4)
\psline[linecolor=black,linewidth=0.5pt]{-}(3,-2.4)(2.5,-3.6)
\psline[linecolor=black,linewidth=0.5pt]{-}(3,-2.4)(3.5,-3.6)
\psline[linecolor=black,linewidth=0.5pt]{-}(1,-2.4)(0.5,-3.6)
\psline[linecolor=black,linewidth=0.5pt]{-}(1,-2.4)(1.5,-3.6)

\begin{scriptsize}
\uput{0}[280](-0.65,-0.4){$\mathsf{ L_1}$}
\uput{0}[280](1.7,-0.4){$\mathsf {R_1}$}
\uput{0}[280](1.3,-1.6){$\mathsf{l}$}
\uput{0}[280](2.6,-1.6){$\mathsf{r}$}
\uput{0}[280](2.55,-2.8){$\mathsf{L_3}$}
\uput{0}[280](3.4,-2.8){$\mathsf{R_3}$}
\uput{0}[280](0.55,-2.8){$\mathsf{L_2}$}
\uput{0}[280](1.4,-2.8){$\mathsf{R_2}$}
\end{scriptsize}

\begin{scriptsize}
\uput{0}[280](0.5,0.3){$\mathbf{1.1}$}
\uput{0}[280](2.1,-0.85){$\mathbf{2.1}$}
\uput{0}[280](3.1,-2.05){$\mathbf{1.3}$}
\uput{0}[280](0.8,-2.05){$\mathbf{1.2}$}
\end{scriptsize}

\begin{scriptsize}
\uput{0}[280](-1.05,-1.35){$2$\mbox{ , }$4$}
\uput{0}[280](0.45,-3.75){$3$\mbox{ , }$1$}
\uput{0}[280](1.45,-3.75){$3$\mbox{ , }$3$}
\uput{0}[280](2.45,-3.75){$2$\mbox{ , }$1$}
\uput{0}[280](3.45,-3.75){$4$\mbox{ , }$2$}
\end{scriptsize}

}
\end{pspicture}

%\end{document}
\vspace{-1.4cm}
\caption{Example of two--agent perfect--information extensive--form game, $\mathbf x.\mathbf y$ denote the $\mathbf y$--th node of agent~$\mathbf x$.}
\label{fig:example}
\vspace{-0.2cm}
\end{figure}
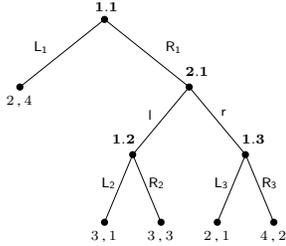

\textbf{(Reduced) Normal form~\cite{neumann1944}}. It is a tabular representation in which each normal--form action, called \emph{plan} and denoted by $p\in P_i$ where $P_i$ is the set of plans of agent~$i$, specifies one action $a$ per information set. We denote by $\boldsymbol \pi_i$ a normal--form strategy of agent~$i$ and by $\pi_{i}(p)$ the probability associated with plan~$p$. The number of plans (and therefore the size of the normal form) is exponential in the size of the game tree. The \emph{reduced normal form} is obtained from the normal form by deleting replicated strategies~\cite{reducednormalform}. Although reduced normal form can be much smaller than normal form, it is exponential in the size of the game tree.

\vspace{-0.2cm}

\begin{exam}\label{exam:1}
The reduced normal form of the game in Fig.~\ref{fig:example} and a pair of normal--form strategies are:

\vspace{-0.5cm}
\begin{scriptsize}
\[
\begin{array}{rc|c|c|}
					\multicolumn{4}{r}{\textnormal{agent~2}} 			\\
									&							& \mathsf{l}		& \mathsf{r}	 	\\ \cline{2-4}
\multirow{5}{*}{\begin{sideways}agent 1\end{sideways}}		&\mathsf{L_1*}					& 2,4				& 2,4				\\ \cline{2-4}
									&\mathsf{R_1L_2L_3}			& 3,1				& 2,1				\\ \cline{2-4}
									&\mathsf{R_1L_2R_3}			& 3,1				& 4,2				\\ \cline{2-4}
									&\mathsf{R_1R_2L_3}			& 3,3				& 2,1				\\ \cline{2-4}
									&\mathsf{R_1R_2R_3}			& 3,3				& 4,2				\\ \cline{2-4}
									\end{array}
~
\pi_{1}=
\begin{cases}
\pi_{1,\mathsf{L_1*}} & = \frac{1}{3}\\ 
\pi_{1,\mathsf{R_1L_2L_3}} & = 0\\
\pi_{1,\mathsf{R_1L_2R_3}} & = \frac{1}{3}\\
\pi_{1,\mathsf{R_1R_2L_3}} & = 0\\
\pi_{1,\mathsf{R_1R_2R_3}} & = \frac{1}{3}
\end{cases}
~
\pi_{2}=
\begin{cases}
\pi_{2,\mathsf{l}} & = 1\\ 
\pi_{2,\mathsf{r}} & = 0\\
\end{cases}
\]
\end{scriptsize}
\vspace{-0.3cm}

\end{exam}
\textbf{Agent form~\cite{kuhn1950,selten1975}}. It is a tabular representation in which each agent is replicated in a number of fictitious agents, each per information set, and all the fictitious agents of the same agent have the same utility. A strategy  is commonly said \emph{behavioral} and denoted by $\boldsymbol \sigma_i$. We denote by $\sigma_{i}(a)$ the probability associated with action~$a\in A_i$. The agent form is linear in the size of the game tree. 
%\textcolor{red}{L'ESEMPIO DI AGENT FORM LO LASCIAMO? SE SI' LO SISTEMO}
%
%\begin{scriptsize}
%\[
%%\small
%\begin{array}{rc|c|c|}
%					\multicolumn{4}{r}{\textnormal{agent~2.1}} 			\\
%									&						& \mathsf{l_1}		& \mathsf{r_1}		 \\ \cline{2-4}
%\multirow{2}{*}{\textnormal{agent~1.1}}		&\mathsf{L_1}				& 1,1				& 1,1				\\ \cline{2-4}
%									&\mathsf{R_1}				& 1,1				& 1,1				\\ \cline{2-4}\\
%					&\multicolumn{3}{c}{\mathsf{L_2}} 			\\
%									\end{array}
%\qquad\qquad
%\begin{array}{rc|c|c|}
%					\multicolumn{4}{r}{\textnormal{agent~2.1}} 			\\
%									&						&\mathsf{ l_1}		&\mathsf{ r_1}		\\ \cline{2-4}
%\multirow{2}{*}{\textnormal{agent~1.1}}		&\mathsf{L_1}				& 1,1				& 1,1				\\ \cline{2-4}
%									&\mathsf{R_1}				& 1,1				& 0,0				\\ \cline{2-4}\\
%					&\multicolumn{3}{c}{\mathsf{R_2}} 			\\
%									\end{array}
%\]
%\vspace{-0.3cm}
%\[
%\qquad\qquad\textnormal{agent~1.2}
%\]
%\end{scriptsize}
%where we assign each fictitious agent the label of the information set in which the fictitious agent plays. An example of behavioral strategies is:
%\begin{scriptsize}
%\[
%%\small
%\sigma_{1}=
%\begin{cases}
%\sigma_{1,\mathsf{L_1}} & = 1.0\\ 
%\sigma_{1,\mathsf{R_1}} & = 0.0\\
%\sigma_{1,\mathsf{L_2}} & = 0.0\\
%\sigma_{1,\mathsf{R_2}} & = 1.0
%\end{cases}
%\quad
%\sigma_{2}=
%\begin{cases}
%\sigma_{2,\mathsf{l_1}} & = 0.8\\ 
%\sigma_{2,\mathsf{r_1}} & = 0.2\\
%\end{cases}
%\]
%\end{scriptsize}

\textbf{Sequence form~\cite{stengel1996}}.
It is a representation constituted by a tabular and a set of constraints. Sequence--form actions are called \emph{sequences}. A sequence $q\in Q_i$ of agent~$i$ is a set of consecutive actions $a\in A_i$ where $Q_i\subseteq Q$ is the set of sequences of agent~$i$ and $Q$ is the set of all the sequences. A sequence can be \emph{terminal}, if, combined with some sequence of the opponents, it leads to a terminal node, or \emph{non--terminal} otherwise. In addition, the initial sequence of every agent, denoted by~$\mathsf{q_\emptyset}$,  is said \emph{empty sequence} and, given sequence~$q\in Q_i$ leading to some information set~$h\in H_i$, we say that~$q'$ \emph{extends}~$q$ and we denote by $q'=q|a$ if the last action of~$q'$ (denoted by $a(q')=a'$) is some action~$a\in \rho(h)$ and $q$ leads to~$h$. We denote by $w=h(q)$ the node $w$ with $a(q)\in\rho(w)$; by $q'\subseteq q$ a subsequence of $q$;  by $\mathbf{x}_i$ the sequence--form strategy of agent~$i$ and by $x_{i}(q)$ the probability associated with sequence~$q\in Q_i$. Finally, condition $q\rightarrow h$ is true if sequence~$q$ crosses information set~$h$. Well--defined strategies are such that, for every information set~$h\in H_i$, the probability~$x_{i}(q)$ assigned to the sequence~$q$ leading to~$h$ is equal to the sum of the probabilities~$x_{i}(q')$s where $q'$ extends~$q$ at~$h$. Sequence form constraints are $x_{i}(\mathsf{q_{\emptyset}})=1$ and $x_{i}(q)=\sum_{a\in\rho(w)}x_{i}(q|a)$ for every sequence~$q$, action~$a$, node~$w$ such that $w=h(q|a)$, and for every agent~$i$.
%can be conveniently described as $F_i \cdot \mathbf{x}_i = \mathbf{f}_i$, where $F_i$ is an opportune matrix and $\mathbf{f}_i$ is an opportune vector, both with entries in~$\{-1,0,1\}$.
The agent~$i$'s utility is represented as a sparse multi--dimensional array, denoted, with an abuse of notation, by $U_i$, specifying the value associated with every combination of terminal sequences of all the agents. The size of the sequence form  is linear in the size of the game tree. 
%Since in this paper we frequently exploit the correspondence between actions and sequences, we denote by~$a(q)$ the last action of sequence~$q$. 
%\textcolor{blue}{[DEFINIREI LA NOTAZIONE $q\subseteq q'$]}

\vspace{-0.2cm}

\begin{exam}\label{exam:2}
The sequence form of the game in Fig.~\ref{fig:example} and a pair of sequence--form strategies are:

\begin{scriptsize}
\vspace{-0.4cm}
\[
%\small
\begin{array}{rc|c|c|c|}
	\multicolumn{2}{c}{}	&			\multicolumn{3}{c}{\textnormal{agent~2}} 			\\
									&						& \mathsf{q_\emptyset}	& \mathsf{l}		& \mathsf{r}		 \\ \cline{2-5}
\multirow{7}{*}{\begin{sideways}agent 1\end{sideways}}		& \mathsf{q_\emptyset}		&					& 				& 				\\ \cline{2-5}
									& \mathsf{L_1}				& 2,4					& 				& 				\\ \cline{2-5}
									& \mathsf{R_1}				&					& 				& 				\\ \cline{2-5}
									&\mathsf{ R_1L_2}			& 					& 3,1				& 				\\ \cline{2-5}
									&\mathsf{ R_1R_2}			&					& 3,3				& 				\\ \cline{2-5}
									&\mathsf{ R_1L_3}			& 					& 				& 2,1				\\ \cline{2-5}
									&\mathsf{ R_1R_3}			&					& 				& 4,2				\\ \cline{2-5}
									\end{array}
\quad 
\mathbf{x}_1 =
\left[
\begin{array}{c}
1 \\ \frac{1}{3} \\ \frac{2}{3} \\ \frac{1}{3} \\\frac{1}{3} \\0 \\\frac{2}{3} 
\end{array}
\right]
\quad
\mathbf{x}_2 =
\left[
\begin{array}{c}
1 \\ 1 \\ 0
\end{array}
\right]
\]
\end{scriptsize}

%\noindent the constraints $F_1\cdot \mathbf{x}_1 = \mathbf{f}_1$ and $F_2\cdot \mathbf{x}_2 = \mathbf{f}_2$ are:
%
%\begin{scriptsize}
%\[
%%\small
%F_1 = 
%\left[
%\begin{array}{ccccccc}
%\textcolor{white}{-}1 	& 0	& \textcolor{white}{-}0	& 0	& 0	& 0	& 0	\\
%-1 	& 1	& \textcolor{white}{-}1	& 0	& 0	& 0	& 0	\\
%0 	& 0	& -1	& 1	& 1	& 0	& 0	\\
%0 	& 0	& -1	& 0	& 0	& 1	& 1	\\\end{array}
%\right], 
%\mathbf{f}_1 = 
%\left[
%\begin{array}{c}
%1 \\ 0 \\ 0\\ 0
%\end{array}
%\right],
%\]
%\[
%F_2 = 
%\left[
%\begin{array}{ccc}
%\textcolor{white}{-}1 	& 0	& 0	\\
%-1 	& 1	& 1	\\
%\end{array}
%\right], 
%\mathbf{f}_2 = 
%\left[
%\begin{array}{c}
%1 \\ 0
%\end{array}
%\right]
%\]
%\end{scriptsize}

\end{exam}

\vspace{-0.4cm}
\textbf{Replicator dynamics}. The standard discrete--time replicator equation with two agents is~\cite{cressman2003}:

\vspace{-0.3cm}
\begin{scriptsize}
\begin{align}\label{eqn:discrep1}
\pi_{1}(p,t+1)&= \pi_{1}(p,t) \cdot  \dfrac{\mathbf e_{p}^{T}\cdot U_{1}\cdot \boldsymbol \pi_{2}(t)}{\boldsymbol \pi_{1}^{T}(t)\cdot U_{1}\cdot \boldsymbol \pi_{2}(t)}\\\label{eqn:discrep2}
\pi_{2}(p,t+1)&= \pi_{2}(p,t) \cdot \dfrac{\boldsymbol \pi_{1}^{T}(t)\cdot U_{2}\cdot \mathbf e_{p}}{\boldsymbol \pi_{1}^{T}(t)\cdot U_{2}\cdot \boldsymbol \pi_{2}(t)}
\end{align}
\end{scriptsize}
\vspace{-0.35cm}

\noindent  while the continuous--time one is 

\vspace{-0.35cm}
\begin{scriptsize}
\begin{align}\label{eqn:contrep1}
\dot \pi_{1}(p)&= \pi_{1}(p) \cdot [(\mathbf e_{p}-\boldsymbol \pi_{1})^{T}\cdot U_{1}\cdot \boldsymbol \pi_{2}]\\\label{eqn:contrep2}
\dot \pi_{2}(p)&= \pi_{2}(p) \cdot [\boldsymbol \pi_{1}^{T}\cdot U_{2}\cdot (\mathbf e_{p}-\boldsymbol \pi_{2})]
\end{align}
\end{scriptsize}
\vspace{-0.35cm}

\noindent where $\mathbf e_{p}$ is the vector in which the $p$--th component is ``1'' and the others are ``0''.

\vspace{-0.05cm}

\vspace{-0.2cm}
\section{Discrete--time replicator dynamics for sequence--form representation}

Initially, we show that the standard discrete--time replicator dynamics for normal form cannot be directly applied when sequence form is adopted. Standard replicator dynamics applied to the sequence form is easily obtained by considering each  sequence~$q$ as a plan~$p$ and thus substituting $\mathbf e_{q}$ to $\mathbf e_{p}$ in~(\ref{eqn:discrep1})--(\ref{eqn:discrep2}) where $\mathbf e_{q}$ is zero for all the components $q'$ such that $q'\neq q$ and one for the component $q'$ such that $q'=q$.

\begin{prop}
The replicator (\ref{eqn:discrep1})--(\ref{eqn:discrep2}) does not satisfy the sequence--form constraints.
\end{prop}

\emph{Proof.} The proof  is by counterexample. Consider $\mathbf{x}_1(t)$ and $\mathbf{x}_2(t)$ equal to the strategies used in Example~\ref{exam:2}. At time $t+1$ the strategy profile generated by (\ref{eqn:discrep1})--(\ref{eqn:discrep2}) is:

\vspace{-0.3cm}
\begin{scriptsize}
\[
\mathbf{x}_1^{T}(t+1) =
\left[
\begin{array}{ccccccc}
0 & \frac{1}{3} & 0 & \frac{1}{2} & \frac{1}{6} & 0 & 0
\end{array}
\right]
\quad
\mathbf{x}_2^{T}(t+1) =
\left[
\begin{array}{ccc}
\frac{1}{2} & \frac{1}{2} & 0
\end{array}
\right]
\]
\end{scriptsize}
\vspace{-0.3cm}

\noindent that does not satisfy the sequence--form constraints, e.g., $x_i(\mathsf{q_\emptyset},t+1)\neq 1$ for all~$i$.$\hfill\Box$

The critical issue behind the failure of the standard replicator dynamics lies in the definition of vector $\mathbf e_{q}$. Now we describe how the standard discrete--time replicator dynamics can be modified to be applied to the sequence form. In our variation, we substitute $\mathbf e_{q}$ with an opportune vector $\mathbf g_{q}$ that depends on the strategy $\mathbf{x}_i(t)$ and it is generated as described in Algorithm~\ref{alg:solveQPE2}, obtaining:

\vspace{-0.4cm}
\begin{scriptsize}
\begin{align}
\label{eqn:discrep3}
x_{1}(q,t+1)&= x_{1}(q,t) \cdot  \frac{\mathbf g_{q}^{T}(\mathbf x_{1}(t))\cdot U_{1}\cdot \mathbf x_{2}(t)}{\mathbf x_{1}^{T}(t)\cdot U_{1}\cdot \mathbf x_{2}(t)}\\ 
\label{eqn:discrep4}
x_{2}(q,t+1)&= x_{2}(q,t) \cdot \frac{\mathbf x_{1}^{T}(t)\cdot U_{2}\cdot \mathbf g_{q}(\mathbf x_{2}(t))}{\mathbf x_{1}^{T}(t)\cdot U_{2}\cdot \mathbf x_{2}(t)}
\end{align}
\end{scriptsize}
\vspace{-0.3cm}

%where $\mathbf g_{q}(\mathbf{x}_i(t))$ is an opportune vector generated as described in Algorithm~\ref{alg:solveQPE2} and used to evaluate the fitness of sequence~$q$. Such a vector substitutes vector $\mathbf{e}_q$. 
%

%\textcolor{red}{[ALGORITMO GIUSTO, RILEGGERE E CONTROLLARE]}

\noindent {The basic idea behind the construction of vector $g_q$ is:
\begin{itemize}
\item assigning ``$1$'' to the probability of all the sequences contained in $q$,
\item normalizing the probability of the sequences extending the contained in $q$,
\item assigning ``$0$'' to the probability of all the other sequences.
\end{itemize}}

\noindent We describe the generation of vector~$\mathbf{g}_q(\mathbf x_{i}(t))$, for clarity we use as running example the generation of $\mathbf g_{\mathsf{R_1R_3}}(\mathbf x_{1}(t))$ related to Example~\ref{exam:2}:
\begin{itemize}
\item all the components of $\mathbf{g}_q(\mathbf x_{i}(t))$ are initialized equal to ``0'', e.g.,

\vspace{-0.3cm}
\begin{scriptsize}
\[
\mathbf g_{\mathsf{R_1R_3}}(\mathbf x_{1}(t))^{T} =
\left[
\begin{array}{ccccccc}
0 & 0 & 0 & 0 & 0 & 0 & 0
\end{array}
\right]
\]
\end{scriptsize}
\vspace{-0.5cm}

\item if sequence~$q$ is played, the algorithm assigns:
\begin{itemize}
\item ``$1$'' to all the components  $g_{q}(q',\mathbf{x}_i(t))$ of $\mathbf g_{q}(\mathbf{x}_i(t))$ where $q'\subseteq q$ (i.e., $q'$ is a subsequence of $q$), e.g.,

\vspace{-0.25cm}
\begin{scriptsize}
\[
\mathbf g_{\mathsf{R_1R_3}}(\mathbf x_{1}(t))^{T} =
\left[
\begin{array}{ccccccc}
1 & 0 & 1 & 0 & 0 & 0 & 1
\end{array}
\right]
\]
\end{scriptsize}
\vspace{-0.3cm}

\item ``$\frac{x_{i}(q'',t)}{x_{i}(q',t)}$'' to all the components $g_{q}(q'',\mathbf{x}_i(t))$ of $\mathbf g_{q}(\mathbf{x}_i(t))$ where $q'\subseteq q$ with $q'=q''\cap q$ and sequence~$q''$ is defined as $q''=q'|a|\ldots$ with $a\in \rho(h)$ and $q \not \rightarrow h$ (i.e., $q'$ is a subsequence of $q$ and $q''$ extends $q'$ off the path identified by $q$), e.g.,

\vspace{-0.3cm}
\begin{scriptsize}
\[
\mathbf g_{\mathsf{R_1R_3}}(\mathbf x_{1}(t))^{T} =
\left[
\begin{array}{ccccccc}
1 & 0 & 1 & \frac{1}{2} & \frac{1}{2} & 0 & 1
\end{array}
\right]
\]
\end{scriptsize}
\vspace{-0.4cm}

\item all the other components are left equal to ``0'',

\end{itemize}
\item if sequence~$q$ is not played, $\mathbf{g}_q(\mathbf{x}_i(t))$ can be arbitrary, since the $q$--th equation of (\ref{eqn:discrep3})--(\ref{eqn:discrep4}) is always zero given that $x_i(q,t)=0$ for every $t$.
\end{itemize}

\begin{algorithm}[t]
\scriptsize
\begin{algorithmic}[1]
\STATE $\mathbf g_{q}(\mathbf x_{i}(t))=\mathbf0$
\IF {$x_{i}(q,t)\neq 0$}
%\STATE $Q'_{i}=\emptyset$
\FOR {$q'\in Q_{i}$ s.t. $q'\subseteq q$}
\STATE $g_{q}(q',\mathbf x_{i}(t))=1$
%\STATE $Q'_{i}=Q'_{i}\cup\{q'\}$
\FOR {$q''\in Q_{i}$ s.t. $q''\cap q = q'$ \textbf{and} $q''=q'|a|\ldots$ : $a\in \rho(h),q \not \rightarrow h$}% $q'\in Q'_{i},\nexists \overline q'\in Q'_{i},q'\subset \overline q' \subset q''$}
%\IF {$a(q'')\neq a, \forall a\in\rho(w),w\in h(Q'_{i})$}
\STATE $g_{q}(q'',\mathbf x_{i}(t))=\frac{x_{i}(q'',t)}{x_{i}(q',t)}$
%\ENDIF
\ENDFOR
\ENDFOR
\ENDIF
\RETURN $\mathbf g_{q}(\mathbf x_{i}(t))$
\end{algorithmic}
\caption{$\mathsf{generate\_\mathbf g_{q}}(\mathbf x_{i}(t))$}
\label{alg:solveQPE2}
\end{algorithm}

\noindent All the vectors $\mathbf g_{q}(\mathbf x_{1}(t))$ of Example~\ref{exam:2} are:

\vspace{-0.35cm}
\begin{scriptsize}
\[
%\small
\begin{array}{c|c|c|c|c|c|c|c|}
						& \mathbf g_{q_{\emptyset}}	&  \mathbf g_{L1}	&  \mathbf g_{R_1}	&  \mathbf g_{R_1L_{2}}	&  \mathbf g_{R_1R_{2}}	&  \mathbf g_{R_1L_{3}}	&  \mathbf g_{R_1R_{3}}	\\ %\cline{2-5}
 \mathsf{q_\emptyset}		&1						& 1				& 1				& 1					& 1					& 1					& 1					\\ %\cline{2-5}
 \mathsf{L_1}				& \frac{1}{3}				& 1				&0				& 0					& 0					& 0					& 0					\\ %\cline{2-5}
 \mathsf{R_1}				&\frac{2}{3}				& 0				&1				& 1					& 1					& 1					& 1					\\ %\cline{2-5}
\mathsf{ R_1L_2}			& \frac{1}{3}				& 0				& \frac{1}{2}		& 1					& 0					& \frac{1}{2}			& \frac{1}{2}			\\ %\cline{2-5}
\mathsf{ R_1R_2}			&\frac{1}{3}				& 0				&  \frac{1}{2}		& 0					& 1					& \frac{1}{2}			& \frac{1}{2}			\\ %\cline{2-5}
\mathsf{ R_1L_3}			& 0						& 0				& 0				& 0					& 0					& 1					& 0					\\ %\cline{2-5}
\mathsf{ R_1R_3}			&\frac{2}{3}				& 0				&1				& 1					& 1					& 0					& 1					\\ %\cline{2-5}
\end{array}
\]
\end{scriptsize}
\vspace{-0.25cm}

We show that replicator dynamics~(\ref{eqn:discrep3})--(\ref{eqn:discrep4}) do not violate sequence--form constraints. 

\vspace{-0.15cm}

\begin{thm}
Given a well--defined sequence--form strategy profile $(\mathbf{x}_1(t),\mathbf{x}_2(t))$, the output strategy profile $(\mathbf{x}_1(t+1),\mathbf{x}_2(t+1))$ of replicator dynamics (\ref{eqn:discrep3})--(\ref{eqn:discrep4}) satisfies sequence--form constraints.
\end{thm}

\vspace{-0.15cm}

\emph{Proof}. The constraints forced by sequence form are:
\begin{itemize}
\item $x_{i}(\mathsf{q_\emptyset},t)=1$ for every~$i$,
\item  $x_{i}(q,t)=\sum_{a\in\rho(w)}x_{i}(q|a,t)$ for every sequence~$q$, action~$a$, node~$w$ such that $w=h(q|a)$, and for every agent~$i$.
\end{itemize}

\vspace{-0.15cm}

Assume, by hypothesis of the theorem, that the above constraints are satisfied at $t$, we need to prove that constraints

\vspace{-0.4cm}
\begin{scriptsize}
\begin{align}\label{eqn:constraint1}
x_{i}(\mathsf{q_{\emptyset}},t+1)&=1\\
x_{i}(q,t+1)&=\sum_{a\in\rho(w)}x_{i}(q|a,t+1) \label{eqn:constraint}
\end{align}
\end{scriptsize}
\vspace{-0.4cm}

\noindent are satisfied. Constraint~(\ref{eqn:constraint1}) always holds because $\mathbf{g}_{\mathsf{q_{\emptyset}}}(\mathbf{x}_1(t))=\mathbf{x}_1(t)$. We rewrite constraints ($\ref{eqn:constraint}$) as

\vspace{-0.45cm}
\begin{scriptsize}
\begin{multline}\label{eqn:constraint2}
x_{i}(q,t) \cdot  \frac{\mathbf g_{q}^{T}(\mathbf x_{i}(t))\cdot U_{i}\cdot \mathbf x_{-i}(t)}{\mathbf x_{i}^{T}(t)\cdot U_{i}\cdot \mathbf x_{-i}(t)}=\\=\sum_{a\in\rho(w)}\left(x_{i}(q|a,t) \cdot  \frac{\mathbf g_{q|a}^{T}(\mathbf x_{i}(t))\cdot U_{i}\cdot \mathbf x_{-i}(t)}{\mathbf x_{i}^{T}(t)\cdot U_{i}\cdot \mathbf x_{-i}(t)}\right)
\end{multline}
\end{scriptsize}
\vspace{-0.4cm}

\noindent Conditions~(\ref{eqn:constraint2}) hold if the following condition holds
\vspace{-0.4cm}

\begin{scriptsize}
\begin{align}\label{eqn:constraint3}
x_{i}(q,t) \cdot  \mathbf g_{q}^{T}(\mathbf x_{i}(t))=\sum_{a\in\rho(w)}\left(x_{i}(q|a,t) \cdot  \mathbf g_{q|a}^{T}(\mathbf x_{i}(t))\right)
\end{align}
\end{scriptsize}
\vspace{-0.4cm}

\noindent Notice that condition~(\ref{eqn:constraint3}) is a vector of equalities, one per sequence $q'$. Condition~(\ref{eqn:constraint3}) is trivially satisfied for components~$q'$ such that $g_{q}(q',\mathbf x_{i}(t))=0$. To prove the condition for all the other components, we introduce two lemmas. 

%\textcolor{red}{[EQUAZIONE VETTORIALE]}

%\textcolor{blue}{[CREDO CHE SIA PIU' COMPLICATO DA SPIEGARE. CREDO CHE TU VOGLIA DIRE: data un $\overline{q}$ valgono certe proprietˆ', il problema e' che se la q nel lemma e' generica alla q' e' generica ed e' come se dicessi che vale per tutte le sequenze]}

\vspace{-0.2cm}

\begin{lem}\label{lem:sub}
Constraint~(\ref{eqn:constraint3}) holds for all components $g_{q}(q',\mathbf x_{i}(t))$  of $\mathbf g_{q}(\mathbf x_{i}(t))$ such that $q'\subseteq q$.
\end{lem}

\emph{Proof}. By construction, $%g_{q}(q,\mathbf x_{i}(t))=
g_{q}(q',\mathbf x_{i}(t))=1$ for every $q'\subseteq q$. 
For every extension $q|a$ of $q$, we have that $q'\subseteq q\subset q|a$.
%\textcolor{blue}{[NON SI CAPISCE] Every $q'\subseteq q$ is also part of each extension $q|a$ of $q$ 
%$q''$ such that $a=a(q'')$ in the right side of (\ref{eqn:constraint3old}), 
For this reason $g_{q|a}(q',\mathbf x_{i}(t))=1
%, \forall a\in \rho(w)
$. Thus

\vspace{-0.3cm}
\begin{scriptsize}
\begin{align*}
\begin{split}
x_{i}(q,t)\cdot g_{q}(q',\mathbf x_{i}(t))&=\sum_{a\in\rho(w)}\left(x_{i}(q|a,t)\cdot g_{q|a}(q',\mathbf x_{i}(t))\right) \hspace{0.5cm}\text{iff} \\
x_{i}(q,t)\cdot 1&=\sum_{a\in\rho(w)}x_{i}(q|a,t)\cdot 1
\end{split}
\end{align*}
\end{scriptsize}

\vspace{-0.3cm}

\noindent that holds by hypothesis. Therefore the lemma is proved.~\hfill$\Box$

\vspace{-0.15cm}

%\textcolor{blue}{[MENTRE IL LEMMA PRECEDENTE E' FACILMENTE COMPRENSIBILE NELLO STATEMENT, QUESTO E' INCOMPRENSIBILE]}
\begin{lem}\label{lem:sup}
Constraint (\ref{eqn:constraint3}) holds for all components $g_{q}(q'',\mathbf{x}_i(t))$ of $\mathbf g_{q}(\mathbf{x}_i(t))$ where $q'\subseteq q$ with $q'=q''\cap q$ and sequence~$q''=q'|a'|\ldots$ with $a'\in \rho(h)$ and $q \not \rightarrow h$.
%$q'\in Q'_{i},\nexists \overline q'\in Q'_{i},q'\subset \overline q' \subset q''$ and $a(q'')\neq a, \forall a\in\rho(w),w\in h(Q'_{i})$.
\end{lem}

\vspace{-0.15cm}

\emph{Proof}. For all $q''$, $g_{q}(q'',\mathbf x_{i}(t))=\frac{x_{i}(q'',t)}{x_{i}(q',t)}$ by construction. In the right side term of (\ref{eqn:constraint3}), for all $a$ we can have either $q|a\nsubset q''$ or  $q|a\subset q''$. In the former we have that $g_{q|a}(q'',\mathbf x_{i}(t))=\frac{x_{i}(q'',t)}{x_{i}(q',t)}$, in the latter there exists only one action $a$ such that $g_{q|a}(q'',\mathbf x_{i}(t))=\frac{x_{i}(q'',t)}{x_{i}(q|a,t)}$, while for the other actions $a^{*}$ the value of $g_{q|a^{*}}(q'',\mathbf x_{i}(t))$ is zero. Hence, we can have two cases: if $q|a\nsubset q''$, then

\vspace{-0.35cm}

\begin{scriptsize}
\begin{align*}
\begin{split}
x_{i}(q,t)\cdot g_{q}(q'',\mathbf x_{i}(t))&=\sum_{a\in\rho(w)}\left(x_{i}(q|a,t)\cdot g_{q|a}(q'',\mathbf x_{i}(t))\right)\hspace{0.5cm} \text{iff}\\
x_{i}(q,t)\cdot \frac{x_{i}(q'',t)}{x_{i}(q',t)}&=\sum_{a\in\rho(w)}\left(x_{i}(q|a,t)\cdot \frac{x_{i}(q'',t)}{x_{i}(q',t)}\right)
\end{split}
\end{align*}
\end{scriptsize}

\vspace{-0.35cm}

\noindent that holds by hypothesis, otherwise if $q|a\subset q''$, then

\vspace{-0.35cm}
\begin{scriptsize}
\begin{align*}
\begin{split}
x_{i}(q,t)\cdot g_{q}(q'',\mathbf x_{i}(t))&=\sum_{a\in\rho(w)}\left(x_{i}(q|a,t)\cdot g_{q|a}(q'',\mathbf x_{i}(t))\right)\hspace{0.5cm} \text{iff}\\
 x_{i}(q,t)\cdot \frac{x_{i}(q'',t)}{x_{i}(q,t)}&=x_{i}(q|a,t)\cdot \frac{x_{i}(q'',t)}{x_{i}(q|a,t)}
\end{split}
\end{align*}
\end{scriptsize}
\vspace{-0.3cm}

\noindent that always holds. Therefore the lemma is proved.\hfill$\Box$

%\begin{lem}\label{lem:equ}
%Constraint (\ref{eqn:constraint3}) holds for all $q'$ such that $a(q')\in \rho(w),w\in h(Q'_{i})$.
%\end{lem}
%
%\emph{Proof}. For all $q'$, $g_{q}(q',t)$ can be either ``0'', if $q'\nsubset q$ or ``1'', if $q'\subset q$. But if $q'\nsubset q$ then $q'\nsubset q|a$ for all $a$ and if $q'\subset q$ then $q'\subset q|a$ for all $a$.
%Thus,
%\begin{align}
%x_{i}(q,t)\cdot g_{q}(q',t)=\sum_{a\in\rho(w)}x_{i}(q|a,t)\cdot g_{q|a}(q',t)\hspace{1cm} \Leftrightarrow\hspace{1cm}x_{i}(q,t)\cdot 1=\sum_{a\in\rho(w)}x_{i}(q|a,t)\cdot 1
%\end{align}
%that holds by hypothesis, or
%\begin{align}
%x_{i}(q,t)\cdot g_{q}(q',t)=\sum_{a\in\rho(w)}x_{i}(q|a,t)\cdot g_{q|a}(q',t)\hspace{1cm} \Leftrightarrow\hspace{1cm}x_{i}(q,t)\cdot 0=\sum_{a\in\rho(w)}x_{i}(q|a,t)\cdot 0
%\end{align}
%that always holds.
%~\hfill$\Box$

From the application of Lemmas~\ref{lem:sub} and~\ref{lem:sup}, it follows that condition~(\ref{eqn:constraint3}) holds.
% \textcolor{blue}{[DA DIRE CHE LA SEQUENZA VUOTA NON EVOLVE]}
\hfill$\Box$

\vspace{-0.3cm}
\section{Replicator dynamics realization equivalence}

%\begin{defn}[Realization plan]\cite{koller}
%The realization plan of a strategy is defined as the distribution probability induced by the strategy over the terminal nodes.
%\end{defn}
%
There is a well--known relation, based on the concept of \emph{realization}, between normal--form  and sequence--form strategies. In order to exploit it, we introduce two results from~\cite{koller}.

\vspace{-0.1cm}

\begin{defn}[Realization equivalent]
Two strategies of an agent are \emph{realization equivalent} if, for any fixed strategies of the other agents, both strategies define the same probabilities for reaching the nodes of the game tree.
\end{defn}
%\begin{prop}[\cite{koller}]
%Mixed strategies are realization equivalent if and only if they have the same realization plan.
%\end{prop}

\vspace{-0.25cm}

\begin{prop}
For an agent with perfect recall, any normal--form strategy is realization equivalent to a sequence--form strategy.
\end{prop}

\vspace{-0.1cm}

We recall in addition that each pure sequence--form strategy corresponds to a pure normal--form strategy in the reduced normal form~\cite{koller}. 
We can show that the evolutionary dynamics of~(\ref{eqn:discrep3})--(\ref{eqn:discrep4})  are realization equivalent to the evolutionary dynamics of the normal--form replicator dynamics and therefore that the two replicator dynamics evolve in the same way.

Initially, we introduce the following lemma that we will exploit to prove the main result.

\vspace{-0.15cm}

\begin{lem}\label{lem:realization}
Given
\begin{itemize}
\vspace{-0.15cm}
\item  a reduced--normal--form strategy $\boldsymbol \pi_{i}(t)$ of agent~$i$,
\vspace{-0.05cm}
\item  a sequence--form strategy $\mathbf x_{i}(t)$ realization equivalent to $\boldsymbol \pi_{i}(t)$,
\end{itemize} 
\vspace{-0.15cm}
it holds that $x_{i}(q|a,t) \cdot  \mathbf g_{q|a}^{T}(\mathbf x_{i}(t))$ is realization equivalent to $\sum_{p\in P:a\in p}\left(\pi_{i}(p,t) \cdot  \mathbf e_{p}^{T}\right)$ for all $a\in A_{i}$ and $q\in Q_i$ with $q|a\in Q_i$.
\end{lem}

\emph{Proof}. We denote by $\tilde {\mathbf x}_{p}(t)$ the sequence--form strategy realization equivalent to $\mathbf e_{p}(t)$. According to~\cite{koller}, we can rewrite the thesis of the theorem as

\vspace{-0.4cm}

\begin{scriptsize}
\begin{align}
x_{i}(q|a,t) \cdot  {\mathbf g_{q|a}^{T}(\mathbf x_{i}(t))}=\sum_{p\in P:a\in p}\left(\pi_{i}(p,t) \cdot  {\tilde {\mathbf x}_{p}(t)^{T}}\right)&&\forall a\in A_{i}\label{vectorequality100}
\end{align}
\end{scriptsize}

\vspace{-0.35cm}

\noindent Notice that, for each action $a$ and sequence $q$ such that $q|a\in Q_i$, condition~(\ref{vectorequality100}) is a vector of equality conditions. Given $a$ and $q$, two cases are possible:
\begin{enumerate}
\item $x_{i}(q|a,t)=0$ and then $\sum_{p\in P:a\in p}\pi_{i}(p,t)=0$, thus conditions~(\ref{vectorequality100}) hold;
\item $x_{i}(q|a,t)\neq0$, in this case:
\begin{itemize}
\item for all components $g_{q|a}(q',\mathbf x_{i}(t))$  of $\mathbf g_{q|a}(\mathbf x_{i}(t))$ and $\tilde x_{p}(q',t)$ of $\tilde {\mathbf x}_{p}(t)$ such that
$q'\subseteq q|a$, we have that $\tilde {x}_{p}(q',t)=1$ for all $p\in P$ with $a\in p$ and Algorithm~1 sets $g_{q|a}(q',\mathbf x_{i}(t))=1$, thus we can rewrite (\ref{vectorequality100})  as

\vspace{-0.4cm}

\begin{scriptsize}

\begin{align*}
x_{i}(q|a,t) \cdot  { g_{q|a}(q',\mathbf x_{i}(t))}&=\sum_{p\in P:a\in p}\left(\pi_{i}(p,t) \cdot  {\tilde { x}_{p}(q',t)}\right)\hspace{0.5cm}\text{iff} \\
x_{i}(q|a,t) \cdot  1&=\sum_{p\in P:a\in p}\left(\pi_{i}(p,t) \cdot  1\right)
\end{align*}

\end{scriptsize}

\vspace{-0.2cm}

\noindent that holds by hypothesis and thus conditions~(\ref{vectorequality100}) hold;
\item for all components $g_{q|a}(q'',\mathbf x_{i}(t))$  of $\mathbf g_{q|a}(\mathbf x_{i}(t))$ and $\tilde x_{p}(q'',t)$ of $\tilde {\mathbf x}_{p}(t)$ such that
$q''$ such that $q''\cap q = q'$ and sequence~$q''=q'|a'|\ldots$ with $a'\in \rho(h)$ and $q \not \rightarrow h$, we have that $\tilde {x}_{p}(q'',t)=1$ for all $p\in P$ with $a,a(q'')\in p$ and ``0'' otherwise, and Algorithm 1 sets $g_{q|a}(q'',\mathbf x_{i}(t))=\frac{x_{i}(q'',t)}{x_{i}(q',t)}$, thus we can rewrite (\ref{vectorequality100})  as

\vspace{-0.4cm}

\begin{scriptsize}

\begin{align*}
x_{i}(q|a,t) \cdot  { g_{q|a}(q'',\mathbf x_{i}(t))}&=\sum_{p\in P:a\in p}\left(\pi_{i}(p,t) \cdot  {\tilde { x}_{p}(q'',t)}\right)\hspace{0.5cm}\text{iff}\\
x_{i}(q|a,t) \cdot  \frac{x_{i}(q'',t)}{x_{i}(q',t)}&=\sum_{p\in P:a,a(q'')\in p}\left(\pi_{i}(p,t) \cdot  1\right)
\end{align*}

\end{scriptsize}

\vspace{-0.2cm}

Using the relationship with the behavioral strategies, we can write

\vspace{-0.4cm}

\begin{scriptsize}

\begin{align*}
x_{i}(q|a,t) \cdot  \dfrac{x_{i}(q'',t)}{x_{i}(q',t)}=\prod_{a'\in q|a}\sigma_{i}(a',t)\cdot\dfrac{\prod_{a'\in q''}\sigma_{i}(a',t)}{\prod_{a'\in q'}\sigma_{i}(a',t)}
\end{align*}

\end{scriptsize}

\vspace{-0.2cm}

\noindent Being $q'\subseteq q|a$ and $q'\subseteq q''$ we have

\vspace{-0.4cm}

\begin{scriptsize}

\begin{multline*}
x_{i}(q|a,t) \cdot  \frac{x_{i}(q'',t)}{x_{i}(q',t)}=\\
\prod_{a'\in q|a\backslash q'}\sigma_{i}(a',t)\cdot\prod_{a'\in q'}\sigma_{i}(a',t)\cdot\prod_{a'\in q''\backslash q'}\sigma_{i}(a',t)=\\
\prod_{a^{*}\in \bigcup_{a'\in\{a,a(q'')\}}q(a')}\sigma_{i}(a^{*},t)
\end{multline*}

\end{scriptsize}

\vspace{-0.2cm}

\noindent that can be easily rewrite as---for details \cite{appendice}---

\vspace{-0.4cm}

\begin{scriptsize}

\begin{align*}
\sum_{p\in P:a,a(q'')\in p}\pi_{i}(p,t)=\prod_{a^{*}\in \bigcup_{a'\in\{a,a(q'')\}}q(a')}\sigma_{i}(a^{*},t)
\end{align*}

\end{scriptsize}

\vspace{-0.2cm}

\noindent and therefore conditions~(\ref{vectorequality100}) hold.
\end{itemize}
\end{enumerate}
This completes the proof of the lemma.\hfill$\Box$

Now we state the main result. {It allows us to study the evolution of a strategy in a game directly in sequence form, instead of using the normal form, and it guarantees that the two dynamics (sequence and normal) are equivalent.}

\vspace{-0.15cm}

\begin{thm}
Given
\vspace{-0.15cm}
\begin{itemize}
\item a normal--form strategy profile $(\boldsymbol\pi_1(t),\boldsymbol\pi_2(t))$ and its evolution $(\boldsymbol\pi_1(t+1),\boldsymbol\pi_2(t+1))$ according to~(\ref{eqn:discrep1})--(\ref{eqn:discrep2}), 
\item a sequence--form strategy profile $(\mathbf{x}_1(t),\mathbf{x}_2(t))$ and its evolution $(\mathbf{x}_1(t+1),\mathbf{x}_2(t+1))$ according to~(\ref{eqn:discrep3})--(\ref{eqn:discrep4}), 
\end{itemize}
if $(\boldsymbol\pi_1(t),\boldsymbol\pi_2(t))$ and $(\mathbf{x}_1(t),\mathbf{x}_2(t))$ are realization equivalent, then also $(\boldsymbol\pi_1(t+1),\boldsymbol\pi_2(t+1))$ and $(\mathbf{x}_1(t+1),\mathbf{x}_2(t+1))$ are realization equivalent.
\end{thm}
\emph{Proof}. Assume, by hypothesis of the theorem, that $(\mathbf{x}_1(t),\mathbf{x}_2(t))$ is realization equivalent to $(\boldsymbol\pi_1(t),\boldsymbol\pi_2(t))$. Thus, according to~\cite{koller}, for every agent~$i$ it holds

\vspace{-0.4cm}

\begin{scriptsize}
\begin{align*}
x_{i}(q|a,t)=\sum_{p\in P:a\in p} \pi_{i}(p,t) && \forall a\in A_{i}
\end{align*}
\end{scriptsize}

\vspace{-0.2cm}

We need to prove that the following conditions hold:

\vspace{-0.3cm}

\begin{scriptsize}
\begin{align}
x_{i}(q|a,t+1)&=\sum_{p\in P:a\in p} \pi_{i}(p,t+1) && \forall a\in A_{i} \label{cond10}
\end{align}
\end{scriptsize}

\vspace{-0.2cm}

\noindent By applying the definition of replicator dynamics, we can rewrite the conditions~(\ref{cond10}) as:

\vspace{-0.4cm}

\begin{scriptsize}

\begin{multline}
x_{i}(q|a,t) \cdot  \frac{\mathbf g_{q|a}^{T}(\mathbf x_{i}(t))\cdot U_{i}\cdot \mathbf x_{-i}(t)}{\mathbf x_{i}^{T}(t)\cdot U_{i}\cdot \mathbf x_{-i}(t)}=\\=\sum_{p\in P:a\in p}\left(\pi_{i}(p,t) \cdot  \frac{\mathbf e_{p}^{T}\cdot U_{i}\cdot \boldsymbol \pi_{-i}(t)}{\boldsymbol \pi_{i}^{T}(t)\cdot U_{i}\cdot \boldsymbol \pi_{-i}(t)}\right) \qquad \forall a\in A_{i}	\label{cond11}
\end{multline}

\end{scriptsize}

%\vspace{-0.2cm}

\noindent Given that, by hypothesis, $\mathbf x_{i}^{T}(t)\cdot U_{i}\cdot \mathbf x_{-i}(t)=\boldsymbol \pi_{i}^{T}(t)\cdot U_{i}\cdot \boldsymbol \pi_{-i}(t)$, we can rewrite conditions~(\ref{cond11}) as:

\vspace{-0.4cm}

\begin{scriptsize}
\begin{multline*}
x_{i}(q|a,t) \cdot  {\mathbf g_{q|a}^{T}(\mathbf x_{i}(t))\cdot U_{i}\cdot \mathbf x_{-i}(t)}=\\=\sum_{p\in P:a\in p}\left(\pi_{i}(p,t) \cdot  {\mathbf e_{p}^{T}\cdot U_{i}\cdot \boldsymbol \pi_{-i}(t)}\right) \qquad \forall a\in A_{i}
\end{multline*}
\end{scriptsize}
\vspace{-0.2cm}

\noindent These conditions hold if and only if $\sum_{p\in P:a\in p}\left(\pi_{i}(p,t) \cdot  \mathbf e_{p}^{T}\right)$ is realization equivalent to $x_{i}(q|a,t) \cdot  \mathbf g_{q|a}^{T}(\mathbf x_{i}(t))$. By Lemma~\ref{lem:realization}, this equivalence holds.\hfill$\Box$

\vspace{-0.2cm}

\section{Continuous--time replicator dynamics for sequence--form representation}

%\textcolor{red}{[METTERE LA DIPENDENZA ESPLICITA DAL TEMPO NELLE SEGUENTI EQUAZIONI?]}

The sequence--form continuous--time replicator equation is

\vspace{-0.35cm}
\begin{scriptsize}
\begin{align}\label{eqn:contrepl1}
\dot x_{1}(q,t)&= x_{1}(q,t) \cdot [(\mathbf g_{q}(\mathbf x_{1}(t))-\mathbf x_{1}(t))^{T}\cdot U_{1}\cdot \mathbf x_{2}(t)]\\\label{eqn:contrepl2}
\dot x_{2}(q,t)&= x_{2}(q,t) \cdot [\mathbf x_{1}(t)^{T}\cdot U_{2}\cdot (\mathbf g_{q}(\mathbf x_{2}(t)-\mathbf x_{2}(t))]
\end{align}
\end{scriptsize}
\vspace{-0.35cm}

\begin{thm}
Given a well--defined sequence--form strategy profile $(\mathbf{x}_1(t),\mathbf{x}_2(t))$, the output strategy profile $(\mathbf{x}_1(t+\Delta t),\mathbf{x}_2(t+\Delta t))$ of replicator dynamics  (\ref{eqn:contrepl1})--(\ref{eqn:contrepl2}) satisfies sequence--form constraints.
%The replicator dynamics (\ref{eqn:contrepl1})--(\ref{eqn:contrepl2}) is correct with respect to the sequence--form constraints.
\end{thm}
%%%%%%%%%%%%%%%%%%%%%%%%%%%%%%%%%%%%%%%%%%%%%%%%%%%%%%%%%%%%%%%%%%%%%%%%%%%%%%%%%%%%%%%%%
%\textcolor{red}{[QUESTA PROOF E' QUELLA VISTA IERI INSIEME. SECONDO ME SI POSSONO USARE GLI STESSI LEMMI DEL DISCRETO E SALTARE TUTTO A PIE' PARI]}
\emph{Proof}. 
The constraints forced by sequence form are:
\begin{itemize}
\item $x_{i}(\mathsf{q_\emptyset},t)=1$ for every~$i$,
\item  $x_{i}(q,t)=\sum_{a\in\rho(w)}x_{i}(q|a,t)$ for every sequence~$q$, action~$a$, node~$w$ such that $w=h(q|a)$, and for every agent~$i$.
\end{itemize}
Assume, by hypothesis of the theorem, that constraints are satisfied at a given time point $t$, we need to prove that constraints

\vspace{-0.6cm}
\begin{scriptsize}
\begin{align}\label{eqn:constraint1c}
x_{i}(\mathsf{q_{\emptyset}},t+\Delta t)&=1\\
x_{i}(q,t+\Delta t)&=\sum_{a\in\rho(w)}x_{i}(q|a,t+\Delta t) \label{eqn:constraintc}
\end{align}
\end{scriptsize}
\vspace{-0.3cm}

\noindent are satisfied. Constraint~(\ref{eqn:constraint1c}) always holds because $\mathbf{g}_{q_{\empty}}(\mathbf{x}_1(t))=\mathbf{x}_1(t)$. We rewrite constraints ($\ref{eqn:constraintc}$) as

\vspace{-0.35cm}
\begin{scriptsize}
\begin{multline}\label{eqn:constraint2bis}
x_{i}(q,t) \cdot [(\mathbf g_{q}(\mathbf x_{i}(t))-\mathbf x_{i}(t))^{T}\cdot U_{i}\cdot \mathbf x_{-i}(t)]=\\=\sum_{a\in\rho(w)}\left(x_{i}(q|a) \cdot [(\mathbf g_{q|a}(\mathbf x_{i}(t))-\mathbf x_{i}(t))^{T}\cdot U_{i}\cdot \mathbf x_{-i}(t)]\right)
\end{multline}
\end{scriptsize}
\vspace{-0.3cm}

\noindent Conditions~(\ref{eqn:constraint2bis}) hold if the following conditions hold

\vspace{-0.3cm}
\begin{scriptsize}
\begin{align}\label{eqn:constraint3old}
x_{i}(q,t) \cdot  \mathbf g_{q}^{T}(\mathbf x_{i}(t))=\sum_{a\in\rho(w)}\left(x_{i}(q|a,t) \cdot  \mathbf g_{q|a}^{T}(\mathbf x_{i}(t))\right)
\end{align}
\end{scriptsize}
\vspace{-0.3cm}

\noindent Notice that condition~(\ref{eqn:constraint3old}) is a vector of equalities. The above condition is trivially satisfied for components~$q'$ such that $g_{q}(q',\mathbf x_{i}(t))=0$. From the application of Lemmas~\ref{lem:sub} and~\ref{lem:sup}, the condition~(\ref{eqn:constraint3old}) holds also for all the other components. 
\hfill$\Box$

\begin{thm}
Given
\begin{itemize}
\item a normal--form strategy profile $(\boldsymbol\pi_1(t),\boldsymbol\pi_2(t))$ and its evolution $(\boldsymbol\pi_1(t+\Delta t),\boldsymbol\pi_2(t+\Delta t))$ according to~(\ref{eqn:contrep1})--(\ref{eqn:contrep2}), 
\item a sequence--form strategy profile $(\mathbf{x}_1(t),\mathbf{x}_2(t))$ and its evolution $(\mathbf{x}_1(t+\Delta t),\mathbf{x}_2(t+\Delta t))$ according to~(\ref{eqn:contrepl1})--(\ref{eqn:contrepl2}), 
\end{itemize}
if $(\boldsymbol\pi_1(t),\boldsymbol\pi_2(t))$ and $(\mathbf{x}_1(t),\mathbf{x}_2(t))$ are realization equivalent, then also $(\boldsymbol\pi_1(t+\Delta t),\boldsymbol\pi_2(t+\Delta t))$ and $(\mathbf{x}_1(t+\Delta t),\mathbf{x}_2(t+\Delta t))$ are realization equivalent.
\end{thm}
%\begin{thm}
%Given a mixed--strategy in reduced normal form $\boldsymbol \pi(t_{0})$ at time $t_{0}$ and its realization plan $\mathbf x(t_{0})$, the strategy $\mathbf x(t_{0}+\Delta t)$ of sequence--form replicator dynamics (\ref{eqn:contrepl1})--(\ref{eqn:contrepl2}) is the realization plan of $\boldsymbol \pi(t_{0}+\Delta t)$ (strategy of normal--form replicator dynamics (\ref{eqn:contrep1})--(\ref{eqn:contrep2})) for all $\Delta t>0$.
%\end{thm}

\emph{Proof}. 
Assume, by hypothesis of the theorem, that $(\mathbf{x}_1(t),\mathbf{x}_2(t))$ is realization equivalent to $(\boldsymbol\pi_1(t),\boldsymbol\pi_2(t))$. Thus, according to~\cite{koller}, for every agent~$i$ it holds

\vspace{-0.35cm}
\begin{scriptsize}
\begin{align*}
x_{i}(q|a,t)=\sum_{p\in P:a\in p} \pi_{i}(p,t) && \forall a\in A_{i}
\end{align*}
\end{scriptsize}
\vspace{-0.3cm}

We need to prove that the following conditions hold:

\vspace{-0.35cm}
\begin{scriptsize}
\begin{align}
x_{i}(q|a,t+\Delta t)&=\sum_{p\in P:a\in p} \pi_{i}(p,t+\Delta t) && \forall a\in A_{i} \label{cond10c}
\end{align}
\end{scriptsize}
\vspace{-0.3cm}

\noindent By applying the definition of replicator dynamics, we can rewrite the conditions~(\ref{cond10c}) as:

\vspace{-0.4cm}
\begin{scriptsize}
\begin{multline}
x_{i}(q|a,t) \cdot [(\mathbf g_{q|a}(\mathbf x_{i}(t))-\mathbf x_{i}(t))^{T}\cdot U_{i}\cdot \mathbf x_{-i}(t)]=\\=\sum_{p\in P:a\in p}\left(\pi_{i}(p,t) \cdot [(\mathbf e_{p}-\boldsymbol \pi_{i}(t))^{T}\cdot U_{i}\cdot \boldsymbol \pi_{-i}(t)]\right)\qquad \forall a\in A_{i}\label{cond11c}
%x_{i}(q|a,t) \cdot  \frac{\mathbf g_{q|a}^{T}(\mathbf x_{i}(t))\cdot U_{i}\cdot \mathbf x_{-i}(t)}{\mathbf x_{i}^{T}(t)\cdot U_{i}\cdot \mathbf x_{-i}(t)}=\\=\sum_{p\in P:a\in p}\left(\pi_{i}(p,t) \cdot  \frac{\mathbf e_{p}^{T}\cdot U_{i}\cdot \boldsymbol \pi_{-i}(t)}{\boldsymbol \pi_{i}^{T}(t)\cdot U_{i}\cdot \boldsymbol \pi_{-i}(t)}\right) \qquad \forall a\in A_{i}	\label{cond11}
\end{multline}
\end{scriptsize}
\vspace{-0.3cm}

\noindent Given that, by hypothesis, $\mathbf x_{i}^{T}(t)\cdot U_{i}\cdot \mathbf x_{-i}(t)=\boldsymbol \pi_{i}^{T}(t)\cdot U_{i}\cdot \boldsymbol \pi_{-i}(t)$, we can rewrite conditions~(\ref{cond11c}) as:

\vspace{-0.4cm}
\begin{scriptsize}
\begin{multline*}
x_{i}(q|a,t) \cdot  {\mathbf g_{q|a}^{T}(\mathbf x_{i}(t))\cdot U_{i}\cdot \mathbf x_{-i}(t)}=\\=\sum_{p\in P:a\in p}\left(\pi_{i}(p,t) \cdot  {\mathbf e_{p}^{T}\cdot U_{i}\cdot \boldsymbol \pi_{-i}(t)}\right) \qquad \forall a\in A_{i}
\end{multline*}
\end{scriptsize}
\vspace{-0.3cm}

\noindent These conditions hold if and only if $\sum_{p\in P:a\in p}\left(\pi_{i}(p,t) \cdot  \mathbf e_{p}^{T}\right)$ is realization equivalent to $x_{i}(q|a,t) \cdot  \mathbf g_{q|a}^{T}(\mathbf x_{i}(t))$. By Lemma~\ref{lem:realization}, this equivalence holds.\hfill$\Box$

\vspace{-0.2cm}

\section{Analyzing the stability of a strategy profile}

%1) obiettivo (questo c'): voglio studiare la natura di un punto (profilo di strategie)
%2) cosa devo fare per studiare questa cosa (questo c'): usare lo Jacobiano
%c' parzialmente, va detto che devi studiare il segno degli autovalori dello Jacobiano, questo forse non c'
%senza dire come deve essere il segno per avere stabilitˆ, ma rimandando questa cosa a qualche referenza
%altro punto: derivabilitˆ di g, che implica: sia la derivabilitˆ dove  giˆ definita che dove non  definita perch  arbitraria
%quindi, bisogna dire che i termini di $g_q$ definiti (quelli non arbitrari) nell'algoritmo di generazione sono derifabili
%poi bisogna dire che dove l'algoritmo dice che  arbitrario, deve essere definito con le best response
%poi, bisogna dire che le best response potrebbero essere molteplici (ottimo sarebbe un esempio di due righe)
%e quindi che a seconda del sottospazio in cui si trovano le dinamiche si hanno replciatori diversi, dove la differenza risiede nella definizione di $g_q$
%a quel punto dici che, in linea di principio, lo studio della natura di un punto richiederebbe lo studio di molti replicatori (potenzialmente un numero combinatorio) per catturare tutte le combinazioni di best response
%infine, dici: possiamo far vedere che la natura  indipendente dal sottospazio da cui arriva, perch il Jacobiano  sempre lo stesso

%\noindent\textcolor{red}{***********************************************}

We focus on characterizing a strategy profile in terms of evolutionary stability. 
When the continuous--time replicator dynamics for normal--form is adopted, evolutionary stability can be analyzed by studying the eigenvalues of the Jacobian in that point~\cite{Arrowsmith1992}---non--positiveness of the eigenvalues is a necessary condition for asymptotical stability, while strict negativeness of the eigenvalues is sufficient. The Jacobian is 

\vspace{-0.3cm}
\begin{scriptsize}
\begin{align*}
J=\left[\begin{array}{ccc}\dfrac{\partial \dot x_{1}(q_{i},t)}{\partial x_{1}(q_{j},t)}&&\dfrac{\partial \dot x_{1}(q_{i},t)}{\partial x_{2}(q_{l},t)}\\
\\
\dfrac{\partial \dot x_{2}(q_{k},t)}{\partial x_{1}(q_{j},t)}&&\dfrac{\partial \dot x_{2}(q_{k},t)}{\partial x_{2}(q_{l},t)}\end{array}\right]&&\begin{split}\forall q_{i},q_{j}\in Q_{1}, \\q_{k},q_{l}\in Q_{2}\end{split}
\end{align*}
\end{scriptsize}
\vspace{-0.3cm}

In order to study the Jacobian of our replicator dynamics, we need to complete the definition of $\mathbf g_{q}(\mathbf x_{i}(t))$. Indeed, we observe that some components of $\mathbf g_{q}(\mathbf x_{i}(t))$ are left arbitrary by Algorithm~1. Exactly, some $q''$ that are related to $q'$ with $x_{i}(q',t)=0$. While it is not necessary to assign values to such components during the evolution of the replicator dynamics, it is necessary when we study the Jacobian. The rationale follows. If $x_{i}(q',t)=0$, then it will remain zero even after $t$. Instead, if, after the dynamics converged to a point, such a point has $x_{i}(q')=0$ for some $q'$, it might be the case that along the dynamics it holds $x_{i}(q')\neq 0$. Thus, in order to define these components of $\mathbf{g}_q(\mathbf{x}_i(t))$, we need to reason backward, assigning the values that they would have in the case such sequence would be played with a probability that goes to zero. In absence of degeneracy, Algorithm~2 addresses this issue assigning a value of ``1'' to a sequence $q''$ if it is the (unique, the game being non--degenerate) best response among the sequences extending $q'$ and ``0'' otherwise, because at the convergence the agents play only the best response sequences. Notice that, in this case, $\mathbf g_{q}(\mathbf x_{i}(t),\mathbf x_{-i}(t))$ depends on both agents' strategies. 

\vspace{-0.25cm}

\begin{algorithm}
\scriptsize
\begin{algorithmic}[1]
\STATE $\mathbf g_{q}(\mathbf x_{i}(t),\mathbf x_{-i}(t))=\mathbf0$
%\STATE $Q'_{i}=\emptyset$
\FOR {$q'\in Q_{i}$ s.t. $q'\subseteq q$}
\STATE $g_{q}(q',\mathbf x_{i}(t),\mathbf x_{-i}(t))=1$
%\STATE $Q'_{i}=Q'_{i}\cup\{q'\}$
\ENDFOR
\FOR {$q''\in Q_{i}$ s.t. $q''\cap q = q'$ \textbf{and} $q''=q'|a|\ldots$ : $a\in \rho(h),q \not \rightarrow h$}% $q'\in Q'_{i},\nexists \overline q'\in Q'_{i},q'\subset \overline q' \subset q''$}
%\IF {$a(q'')\neq a, \forall a\in\rho(w),w\in h(Q'_{i})$}
\IF {$x_{i}(q',t)\neq0$}
\STATE $g_{q}(q'',\mathbf x_{i}(t),\mathbf x_{-i}(t))=\frac{x_{i}(q'',t)}{x_{i}(q',t)}$
\ELSIF{$q''=\text{argmax}_{q^{*}:a(q^{*})\in\rho(h)}\mathbb E[U_{i}(q^{*},\mathbf x_{-i})]$}
\STATE $g_{q}(q'',\mathbf x_{i}(t),\mathbf x_{-i}(t))=1$
%\ENDIF
\ENDIF
\ENDFOR
\RETURN $\mathbf g_{q}(\mathbf x_{i}(t),\mathbf x_{-i}(t))$
\end{algorithmic}
\caption{$\mathsf{generate\_\mathbf g_{q}}(\mathbf x_{i}(t),\mathbf x_{-i}(t))$}
\label{alg:solveQPE}
\end{algorithm}

\vspace{-0.25cm}

Given the above complete definition of $\mathbf g_{q}$, we can observe that all the components of $\mathbf g_{q}(\mathbf x_{i}(t),\mathbf x_{-i}(t))$ generated by Algorithm~2 are differentiable, being ``0'' or ``1'' or ``$\frac{x_{i}(q'',t)}{x_{i}(q',t)}$''. Therefore, we can derive the Jacobian as:

\vspace{-0.3cm}
\begin{scriptsize}
\[
{\dfrac{\partial \dot x_{1}(q_{i},t)}{\partial x_{1}(q_{j},t)}=}
\begin{cases}
\begin{split}&(\mathbf g_{q_i}(\mathbf x_{1}(t),\mathbf x_{2}(t))-\mathbf x_{1}(t))^{T}\cdot U_{1}\cdot \mathbf x_{2}(t)+x_{1}(q_{i},t)\cdot\\&\cdot\left[\left(\frac{{\partial \mathbf g}_{q_{i}}(\mathbf x_{1}(t),\mathbf x_{2}(t))}{\partial x_{1}(q_{j},t)}-\mathbf e_{i}\right)^{T}\cdot U_{1}\cdot \mathbf x_{2}(t)\right]\end{split}&\text{if $i=j$}\\
x_{1}(q_{i},t)\cdot\left[\left(\dfrac{{\partial \mathbf g}_{q_{i}}(\mathbf x_{1}(t),\mathbf x_{2}(t))}{\partial x_{1}(q_{j},t)}-\mathbf e_{j}\right)^{T}\cdot U_{1}\cdot \mathbf x_{2}(t)\right]&\text{if $i\neq j$}
\end{cases}
\]
\vspace{-0.1cm}
\begin{align*}
%&\dfrac{\partial \dot x_{i}}{\partial x_{j}}=\begin{cases}
%(\mathbf e_{i}-\mathbf x)^{T}\cdot A\cdot \mathbf y-x_{i}\cdot[(\mathbf e_{i}-\mathbf e_{n})^{T}\cdot A\cdot \mathbf y]&\text{if $i=j$}\\
%-x_{i}\cdot[(\mathbf e_{j}-\mathbf e_{n})^{T}\cdot A\cdot \mathbf y]&\text{if $i\neq j$}
%\end{cases}\\
&\dfrac{\partial \dot x_{1}(q_{i},t)}{\partial x_{2}(q_{l},t)}= x_{1}(q_{i},t)\cdot[(\mathbf g_{q_i}(\mathbf x_{1}(t),\mathbf x_{2}(t))-\mathbf x_{1}(t))^{T}\cdot U_{1}\cdot \mathbf e_{l}]\\
&\dfrac{\partial \dot x_{2}(q_{k},t)}{\partial x_{1}(q_{j},t)}=x_{2}(q_{k},t)\cdot[\mathbf e_{j}^{T}\cdot U_{2}\cdot (\mathbf g_{q_k}(\mathbf x_{2}(t),\mathbf x_{1}(t))-\mathbf x_{2}(t))]
%&\dfrac{\partial \dot y_{k}}{\partial y_{l}}=\begin{cases}
%\mathbf x^{T}\cdot B\cdot (\mathbf e_{k}-\mathbf y)-y\cdot[\mathbf x^{T}\cdot B\cdot (\mathbf e_{k}-\mathbf e_{m})]&\text{if $k=l$}\\
%-y_{k}\cdot[(\mathbf x^{T}\cdot B\cdot (\mathbf e_{l}-\mathbf e_{m})]&\text{if $k\neq l$}
%\end{cases}
\end{align*}
\vspace{-0.1cm}
\[
{\dfrac{\partial \dot x_{2}(q_{k},t)}{\partial x_{2}(q_{l},t)}=}
\begin{cases}
\begin{split}&\mathbf x_{1}(t)^{T}\cdot U_{2}\cdot (\mathbf g_{q_{k}}(\mathbf x_{2}(t),\mathbf x_{1}(t))-\mathbf x_{2}(t))+x_{2}(q_{k},t)\cdot\\&\cdot\left[\mathbf x_{1}(t)^{T}\cdot U_{2}\cdot \left(\frac{{\partial\mathbf g}_{q_{k}}(\mathbf x_{2}(t),\mathbf x_{1}(t))}{\partial x_{2}(q_{l},t)}-\mathbf e_{k}\right)\right]\end{split}&\text{if $k=l$}\\
x_{2}(q_{k},t)\cdot\left[\mathbf x_{1}(t)^{T}\cdot U_{2}\cdot \left(\dfrac{{\partial\mathbf g}_{q_{k}}(\mathbf x_{2}(t),\mathbf x_{1}(t))}{\partial x_{2}(q_{l},t)}-\mathbf e_{l}\right)\right]&\text{if $k\neq l$}
\end{cases}
\]
\end{scriptsize}

\vspace{-0.2cm}

%\textcolor{red}{Correggi le derivate parziali di g rispetto a $x_i$.}
With degenerate games, given a opponent's strategy profile $\mathbf x_{-i}(t)$ and a sequence $q\in Q_{i}$ such that $x_{i}(q,t)=0$, we can have multiple best responses. Consider, e.g., the game in Example~\ref{exam:2}, with $\mathbf x_{1}^{T}(t)=[\begin{array}{ccccccc}1 & 1 & 0 & 0& 0& 0& 0\end{array}]$, $\mathbf x_{2}^{T}(t)=[\begin{array}{ccc}1 & 1 & 0 \end{array}]$ and compute $\mathbf g_{\mathsf{R_1L_3}}(\mathbf x_{1}(t),\mathbf x_{2}(t))$: both sequences $\mathsf{R_1L_2}$ and $\mathsf{R_1R_2}$ are best responses to $\mathbf x_{2}(t)$. Reasoning backward, we have different vectors $\mathbf g_{q}(\mathbf x_{i},\mathbf x_{-i})$ for different dynamics. More precisely, we can partition the strategy space around $(\mathbf x_{i},\mathbf x_{-i})$, associating a different best response with a different subspace and therefore with a different $\mathbf g_{q}(\mathbf x_{i},\mathbf x_{-i})$. Thus, in principle, in order to study the stability of a strategy profile, we would need to compute and analyze all the (potentially combinatory) Jacobians. However, we can show that all these Jacobians are the same and therefore, even in the degenerate case, we can safely study the Jacobian by using a $\mathbf g_{q}(\mathbf x_{i},\mathbf x_{-i})$ as generated by Algorithm~2 except, if there are multiple best responses, Step~7--8 assign ``1'' only to one, randomly chosen, best response.
\begin{thm} \label{thm:proofappendix}Given
\begin{itemize}
\item a specific sequence $q\in Q_{i}$ such that $x_{i}(q,t)=0$,
\item a sequence--form strategy $\mathbf x_{-i}(t)$,
\item a sequence $q'\subseteq q$,
\item the number of sequences $q''$ such that $q''\cap q = q'$ and $q''=q'|a|\ldots$ : $a\in \rho(h),q \not \rightarrow h$ and that are best responses to $\mathbf x_{-i}(t)$ is larger than one,
%that are $\left|Q^{*}\right|>1$, where $Q^{*}=\{q''\text{ s.t. }q''\cap q = q'\text{ and }q''=q'|a|\ldots\text{ : }a\in \rho(h),q \not \rightarrow h\text{ and }q''=\text{argmax}_{q^{*}:a(q^{*})\in\rho(w)}\mathbb E[U_{1}(q^{*},\mathbf x_{2})]\}$
\end{itemize}
the eigenvalues of the Jacobian are independent from which sequence $q''$ is choosen as best--response.
\end{thm}

\vspace{-0.2cm}

\section{Conclusions and future works}

In this paper we developed efficient evolutionary game theory techniques to deal with extensive--form games. We designed, to the best of our knowledge, the first replicator dynamics applicable with the sequence form of an extensive--form game, allowing an exponential reduction of time and space w.r.t. the standard (normal--form) replicator dynamics. Our replicator dynamics is realization equivalent w.r.t. the standard one and therefore these two replicator dynamics evolve in the same way. We show the equivalence for both the discrete and continuous time cases. Finally, we discuss how standard tools from dynamical systems for the study of the stability of strategies can be adopted with our continuous--time replicator dynamics.

In future, we intend to explore the following problems: extending the results on multi--agent learning when sequence form is adopted taking into account also Nash refinements for extensive--form games (we recall, while this is possible with sequence form, it is not with the normal form); extending our results to other forms of dynamics, e.g., best response dynamics, imitation dynamics, smoothed best replies, the Brown--von Neumann--Nash dynamics; comparing the expressivity and the effectiveness of replicator dynamics when applied to the three representation forms.

%\appendix
\newpage

\section{Appendix}

\section{Relation between normal--form/behavioral/sequence--form strategies}

We briefly review how realization equivalent strategies can be derived according to~\cite{stengel1996}. 

Given a behavioral strategy $\boldsymbol \sigma_{i}$, we can derive the (realization) equivalent normal--form strategy and sequence--form strategy as follow

\vspace{-0.2cm}
\begin{scriptsize}
\begin{align}\label{eqn:pisigma}
\pi_{i}(p)	&	=\prod_{a\in p:p\in P}\sigma_{i}(a)	& \forall p\in P_i	\\
x_{i}(q)	&	=\prod_{a\in q:q\in Q}\sigma_{i}(a)	& \forall q\in Q_i
\end{align}
\end{scriptsize}
\vspace{-0.2cm}

\noindent Given a normal--form strategy $\boldsymbol \pi_{i}$, we can derive the (realization) equivalent behavioral strategy:

\vspace{-0.2cm}
\begin{scriptsize}
\begin{align}
\sigma_{i}(a)=\sum_{p\in P:a\in p}\pi_{i}(p)
\end{align}
\end{scriptsize}
\vspace{-0.2cm}

\noindent Given a normal--form strategy $\boldsymbol \pi_i$ in \emph{reduced} normal form, we can derive the (realization) equivalent sequence--form strategy:

\vspace{-0.2cm}
\begin{scriptsize}
\begin{align}
x_{i}(q|a)=\sum_{p\in P:a\in p}\pi_{i}(p)
\end{align}
\end{scriptsize}
\vspace{-0.2cm}

\noindent We denote by $q(a)$ the sequence whose last action is $a$. We state the following lemma that we use to prove a main result.
\begin{lem}\label{lem:equivalence}
Given:
\begin{itemize}
\item a normal--form strategy $\boldsymbol \pi_{i}$ in reduced normal form, 
\item its equivalent behavioral strategy $\boldsymbol \sigma_{i}$, 
\item a subset of actions $\{a_{1},\ldots,a_{m}\}\subseteq A_{i}$, 
\end{itemize}
it holds

\vspace{-0.2cm}
\begin{scriptsize}
\begin{align}
\sum_{p\in P:a_{1},\ldots,a_{m}\in p}\pi_{i}(p)=\prod_{a\in \bigcup_{j=1}^{m}q(a_{j})}\sigma_{i}(a)
\end{align}
\end{scriptsize}
\vspace{-0.2cm}

\end{lem}
\emph{Proof}. Suppose that $p=a_{1},\ldots,a_{n}$. By (\ref{eqn:pisigma}) we know that

\vspace{-0.2cm}
\begin{scriptsize}
\begin{align}
\pi_{i}(p)&=\sigma_{i}(a_{1})\cdot \cdots \cdot \sigma_{i}(a_{n})
\end{align}
\end{scriptsize}
\vspace{-0.2cm}

\noindent For all plan of actions $p\in P$ where $\{a_{1},\ldots,a_{m}\}\in p$, given an actiona $a$ such that $a\notin \{a_{1},\ldots,a_{m}\}$,
we can have two possibilities
\begin{enumerate}
%\item $\exists a^{*}\in q(\{a_{1},\ldots,a_{m}\})$ such that $a\in\rho(h)$ where $a^{*}\in\rho(h)$ and $a\neq a^{*}$; in this case for all plan of actions $p'$ if $a^{*}\in p'$ then $a\notin p'$.
\item %$a\in q(\{a_{1},\ldots,a_{m}\})$
$a\in \bigcup_{j=1}^{m}q(a_{j})$, in this case the action $a$ is present in every plan of actions $p$, being always present $\{a_{1},\ldots,a_{m}\}$; thus

\vspace{-0.2cm}
\begin{scriptsize}
\begin{multline*}
\sum_{p\in P:a_{1},\ldots,a_{m}\in p}\pi_{i}(p')=\\ \sigma_{i}(a)\cdot
\prod_{j=1}^{m}\sigma_{i}(a_{j})
\cdot\sum_{p\in P:a_{1},\ldots,a_{m}\in p}\left(\sigma_{i}(a_{m+2})\cdot\cdots\cdot\sigma_{i}(a_{n})\right)
\end{multline*}
\end{scriptsize}
\vspace{-0.2cm}

\item %$ a\notin q(\{a_{1},\ldots,a_{m}\})$
$a\notin \bigcup_{j=1}^{m}q(a_{j})$, in this case there is a subset $P'\subseteq P$ such that there is exactly a $p\in P'$ for each action $a'\in\rho(h)$, where $a\in\rho(h)$. 

%\vspace{-0.2cm}
%\begin{scriptsize}
%\begin{align*}
%\forall p'\in P', \exists a'\in\rho(h), \text{ with $a\in\rho(h)$, } p'=(a_{1},\ldots,a_{m},a',a_{m+2},\ldots,a_{n})
%\end{align*}
%\end{scriptsize}
%\vspace{-0.2cm}

\noindent By definition of behavioral strategy we know that

\vspace{-0.2cm}
\begin{scriptsize}
\begin{align*}
\sum_{a\in\rho(h)}\sigma_{i}(a)=1&&\forall h\in H
\end{align*}
\end{scriptsize}
\vspace{-0.2cm}

\noindent Thus

\vspace{-0.2cm}
\begin{scriptsize}
\begin{multline*}
\hspace{-0.35cm}\sum_{p\in P:a_{1},\ldots,a_{m}\in p}\pi_{i}(p)=\\
\hspace{-0.35cm} \prod_{j=1}^{m}\sigma_{i}(a_{j})\cdot\sum_{p\in P:a_{1},\ldots,a_{m}\in p}\left(\sigma_{i}(a_{m+2})\cdot\cdots\cdot\sigma_{i}(a_{n})\right)\cdot\sum_{a\in\rho(h)}\sigma_{i}(a)=\\
\hspace{-0.35cm} \prod_{j=1}^{m}\sigma_{i}(a_{j})\cdot\sum_{p\in P:a_{1},\ldots,a_{m}\in p}\left(\sigma_{i}(a_{m+2})\cdot\cdots\cdot\sigma_{i}(a_{n})\right)
\end{multline*}
\end{scriptsize}
\vspace{-0.2cm}

\end{enumerate}
Thus, we can write

\vspace{-0.2cm}
\begin{scriptsize}
\begin{align*}
\sum_{p\in P:a_{1},\ldots,a_{m}\in p}\pi_{i}(p)=\sum_{p\in P:a_{1},\ldots,a_{m}\in p}\sigma_{i}(a_{1})\cdot\cdots\cdot\sigma_{i}(a_{n})
\end{align*}
\end{scriptsize}
\vspace{-0.2cm}

\noindent where, by Point 1, we know that all the actions $a$ that are in the path of some $a_{1},\ldots,a_{m}$, $a\in\bigcup_{j=1}^{m}q(a_{j})$, are present in every plan of actions

\vspace{-0.2cm}
\begin{scriptsize}
\begin{multline*}
\sum_{p\in P:a_{1},\ldots,a_{m}\in p}\pi_{i}(p)=\\
=\prod_{a\in \bigcup_{j=1}^{m}q(a_{j})}\sigma_{i}(a)\cdot\sum_{p\in P:a_{1},\ldots,a_{m}\in p}\sigma_{i}(a_{m+k})\cdot\cdots\cdot\sigma_{i}(a_{n})
\end{multline*}
\end{scriptsize}
\vspace{-0.2cm}

\noindent and, by Point 2, the other actions sum to ``1''

\vspace{-0.2cm}
\begin{scriptsize}
\begin{align*}
\sum_{p\in P:a_{1},\ldots,a_{m}\in p}\pi_{i}(p)=\prod_{a\in \bigcup_{j=1}^{m}q(a_{j})}\sigma_{i}(a)
\end{align*}
\end{scriptsize}
\vspace{-0.2cm}

\noindent This completes the proof of the lemma.\hfill$\Box$

\section{Proof of Theorem \ref{thm:proofappendix}}

\emph{Proof}. For each sequence $q''$ that is a best--response to $\mathbf x_{-i}(t)$, we can have different vectors $\mathbf g_{q}(\mathbf x_{i}(t),\mathbf x_{-i}(t))$. Suppose to take two different vectors $\mathbf g_{q}(\mathbf x_{i}(t),\mathbf x_{-i}(t))$ and $\mathbf g'_{q}(\mathbf x_{i}(t),\mathbf x_{-i}(t))$. To prove the equality of the two Jacobians we have to prove that each term is the same. All the terms multiplied by $\mathbf x_{i}(q,t)=0$ can be discarded, they being equal to zero. For this reason the only term different from 0 in the Jacobian is $\frac{\partial\dot x_{i}(q,t)}{\partial\dot x_{i}(q,t)}$, thus we have to prove

\vspace{-0.3cm}
\begin{scriptsize}
\begin{multline}\label{eqn:jacob}
(\mathbf g_{q}(\mathbf x_{i}(t),\mathbf x_{-i}(t))-\mathbf x_{i}(t))^{T}\cdot U_{1}\cdot \mathbf x_{-i}(t)=\\(\mathbf g'_{q}(\mathbf x_{i}(t),\mathbf x_{-i}(t))-\mathbf x_{i}(t))^{T}\cdot U_{i}\cdot \mathbf x_{-i}(t)
\end{multline}
\end{scriptsize}
\vspace{-0.3cm}

\noindent  We can rewrite the equality (\ref{eqn:jacob}) as

\vspace{-0.3cm}
\begin{scriptsize}
\begin{align*}%\label{eqn:jacob2}
\mathbf g_{q}^{T}(\mathbf x_{i}(t),\mathbf x_{-i}(t))\cdot U_{i}\cdot \mathbf x_{-i}(t)=\mathbf {g'}_{q}^{T}(\mathbf x_{i}(t),\mathbf x_{-i}(t))\cdot U_{i}\cdot \mathbf x_{-i}(t)
\end{align*}
\end{scriptsize}
\vspace{-0.3cm}

\noindent that always holds because $\mathbf g_{q}(\mathbf x_{i}(t),\mathbf x_{-i}(t))$ and $\mathbf g'_{q}(\mathbf x_{i}(t),\mathbf x_{-i}(t))$, even if they differ for some components, provide the same expected utility by definition of best response. Even if an agent randomizes over multiple best responses, the theorem holds for the same reason.
\hfill$\Box$

\newpage
\bibliographystyle{alpha}
\bibliography{aaai13}

\end{document}